# Field Dependence of Blocking Temperature in Magnetite Nanoparticles

G. F. Goya[1] and M. P. Morales[2]

[1] Instituto de Física, Universidade de São Paulo, CP 66318, 05315-970 São Paulo, SP Brazil

[2] Instituto de Ciencia de Materiales de Madrid, CSIC. Madrid, Spain.

**Abstract.** Spherical magnetite nanoparticles having average particle size <d> = 5 nm have been synthesized by coprecipitation of Fe(II) and Fe(III) salts in KOH with Polyvinylalcohol (PVA). The resulting dry powder displayed superparamgnetic (SPM) behaviour at room temperature, with a transition to a blocked state at $T_B$ ~ 45 K for applied field $H_{app}$ = 500 Oe. The effect of dipolar interactions was investigated by measuring the dependence of $T_B$ on the applied field $H_{ap}$ and driven ac field in susceptibility data. A thermally activated model has been used to fit the dynamic data to obtain the single-particle energy barriers $E_a = K_{eff}V$, allowing us to estimate the contributions of dipolar interactions to the single-particle effective magnetic anisotropy $K_{eff}$. We have measured the dependence of $T_B$ with $H_{ap}$ in order to draw the transition contours of a H-T diagram. Two different regimes are found for the $(T_B-T_0) \sim H^\lambda$ dependence at low and high fields, that can be understood within a pure SPM relaxation-time (Néel-Brown) landscape. The $T_B(H)$ data shows a crossover from $\lambda = 2/3$ to $\lambda \sim 2$ for applied magnetic fields of $\approx$ 550 Oe.

**Introduction.** Along the last two decades we have witnessed considerable achievements related to the design of magnetic nanostructures, aiming both basic research and technological applications. [1,2] These genuine improvements of fabrication techniques have not been accompanied by a detailed comprehension of the relationships between structural and magnetic properties in fine particle systems, even those composed of 'simple' materials whose bulk properties are well understood. One of the oldest magnetic materials known, magnetite, has been revisited in recent years since it turned out that the magnetic and spin structures in this material are appealing for applications in magnetoelectronic and spin-valve devices. Magnetite $Fe_3O_4$ is a ferrimagnet ($T_C$=860 K) with a cubic spinel structure. The $Fe^{2+}$ and $Fe^{3+}$ ions occupy the tetrahedral (*A*) and octahedral (*B*) sites, with a formal ionic distribution that can be represented by $(Fe^{3+})_A(Fe^{3+}Fe^{2+})_B O^{2-}_4$ i.e. there is a mixed valence of Fe on the octahedral sublattice. The first-order magnetocrystalline anisotropy constant has a negative value (at room temperature) of $K_1$ = -13.5 kJ/m$^3$. This material undergoes a magnetic transition at the Verwey temperaure $T_V$ = 120 K, which is related to structural changes from cubic to triclinic symmetry, yielding to uniaxial magnetic anisotropy with <0 0 1> easy axis. In this work we report on the magnetic properties of $Fe_3O_4$ nanoparticles with average linear dimensions <d> = 5 nm, in order to study the effect of temperature and external field on the transition to the ordered state.

**Experimental Procedure**

Magnetite particles of around 5 nm were obtained following Lee et al. method [3]. The iron salt mixture (Fe(II) and Fe(III)) was added to 1M K(OH) solution with 1wt.% of Polyvinylalcohol (PVA) at room temperature, and afterwards dried in air. X-ray diffraction (XRD) measurements were performed using a Philips 1710 powder diffractometer using Cu-K$_\alpha$ radiation in the 2θ range from 5 to 70 degrees.

---

[1] Corresponding author. E-mail: goya@macbeth.if.usp.br

| <d>$_{TEM}$ (nm) | <d>$_{XRD}$ (nm) | T (K) | H$_C$ (Oe) | M$_S$ (emu/g) | M$_R$ (emu/g) | Hyperfine Parameters @ T = 4.2 K ||||
|---|---|---|---|---|---|---|---|---|---|
| | | | | | | Site | B$_{hyp}$ (T) | QS (mm/s) | IS (mm/s) | I (%) |
| 4 | 4.3 | 5 | 294(10) | 56.1(2) | 10.1(8) | A | 50.1 | 0.03 | 0.45 | 61 |
| | | 300 | 12(7) | 31.8(2) | 0 | B | 52.5 | -0.06 | 0.48 | 39 |

**Table I. Structural, magnetic and hyperfine parameters for Fe$_3$O$_4$ nanoparticles at different tempreatures.**

Mössbauer spectroscopy (MS) measurements were performed with a conventional constant-acceleration spectrometer in transmission geometry with a $^{57}$Co/Rh suorce between 4.2 and 296 K. Static and dynamic magnetic measurements as a function of frequency and temperature were performed in a commercial SQUID magnetometer (Quantum Design). Zero-field- cooled (ZFC) curves were taken between 5K and 300 K, for different values of applied field H$_{app}$ (2 Oe < H < 70 kOe).

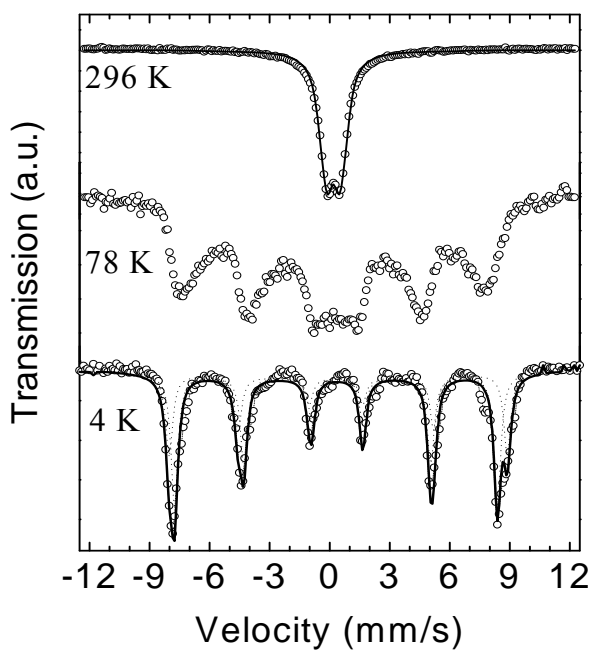

**Figure 1. Mössbauer spectra taken at different temperatures. Solid lines are the best fit to experimental data (open circles).**

**Results and Discussion**

In a previous work [4], we have undertaken a detailed characterization of the present system, and some of the sample data are reproduced for clarity in Table I. The main results can be summarized as follows: samples consisted of single phase magnetite particles with nearly spherical shape and average crystallite size of <d> = 4-5 nm, as seen from X-ray data and TEM images.

Magnetization and Mössbauer data showed superparamagnetic (SPM) behavior at room temperature, and a blocking temperature of $T_B \sim 45$ K. The Mössbauer spectra at T = 78 K displayed relaxation effects, making impossible to obtain the hyperfine parameters (HP's) in the magnetically ordered state. To overcome this, we have recorded the Mössbauer spectrum at 4.2 K as shown in figure 1, obtaining the HP's displayed in Table I. The values of hyperfine field for A and B sites agrees well with previously reported for magnetite at this temperature. No evidence of spin canting has been found through broadening of the inner side (lower velocities) in spectral lines, and moreover the measured spectral linewidth ($\Gamma \sim 0.40$–$0.42$ mm/s) indicates a good crystalline order of the particle cores.

In order to analyze the dynamic magnetic properties, we have measured the temperature dependence of $\chi'(T)$ and $\chi''(T)$ for different frequencies $f$, as shown in figure 2. The data for both components $\chi'(T)$ and $\chi''(T)$ exhibit the expected behavior of SPM systems, i. e., the occurrence of a maximum at a temperature $T_m$ for both $\chi'(T)$ and $\chi''(T)$ components, which shifts towards higher values with increasing frequency. [5]

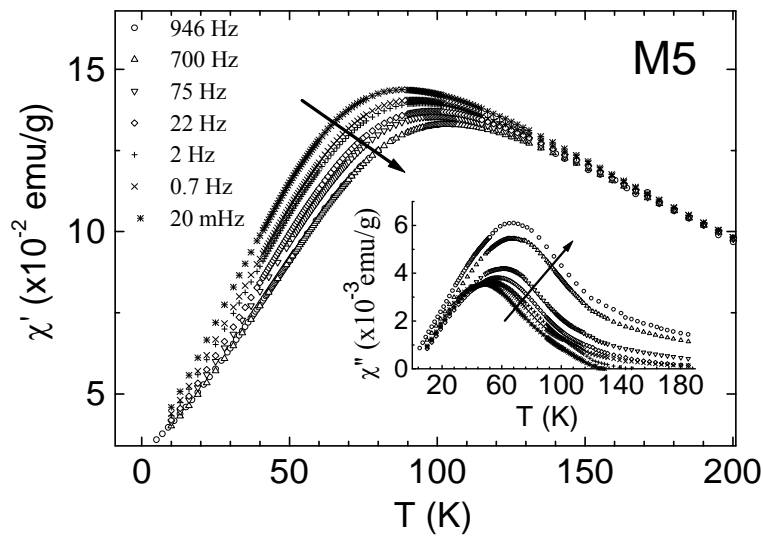

**Figure 2. In-phase component of the ac susceptibility curves $\chi'(T)$ for different applied frequencies. Inset: Out-of-phase component $\chi''(T)$. The arrows indicate increasing frequencies.**

As a useful criterion for classifying the freezing/blocking process observed, we used the empirical parameter $\Phi$, which represents the relative shift of the temperature $T_m$ per decade of frequency. For the present sample, we obtain

$$\Phi = \frac{\Delta T_m}{T_m \Delta \log_{10}(f)} = 0.07$$

Although this value is close to the 0.1-0.13 values found for SPM systems, [5] the slightly inferior value is likely to be originated from interparticle interactions [6], in agreement with the concentrated nature of our samples. The measuring time $\tau_m$ (or frequency $f_m$) of each experimental technique determines the dynamic response of an ensemble of fine particles. As the reversion of the magnetic moment in a single-domain particle over the anisotropy energy barrier $E_a$ is assisted by thermal phonons, the relaxation time τ exhibits an exponential dependence on temperature characterized by a Néel-Brown expression

$$f = f_0 \exp\left(-\frac{E_a}{k_B T}\right) \qquad (1)$$

In the absence of an external magnetic field, the energy barrier can be assumed to be proportional to the particle volume, i.e. $E_a = K_{eff} V$, where $K_{eff}$ is an effective magnetic anisotropy constant. The linear dependence of $\ln(f)$ versus $1/T_B$ observed in fig. 3 indicates that the Néel-Brown model fits the data. Using the average particle radii <d> = 5nm, we obtained an effective anisotropy value of $K_{eff}$ = 356 kJ/m$^3$, which is an order of magnitude larger that the magnetocrystalline anisotropy constant of bulk magnetite $K_1^{bulk}$ = 13.5 kJ/m$^3$, indicating an additional source of magnetic anisotropy to the single-particle energy barriers. Large values of $E_a$ (and $K_{eff}$) are customarily found in particulate systems, which have been associated to dipolar interactions (for concentrated systems) and/or surface effects (mainly in diluted systems). [6,7]

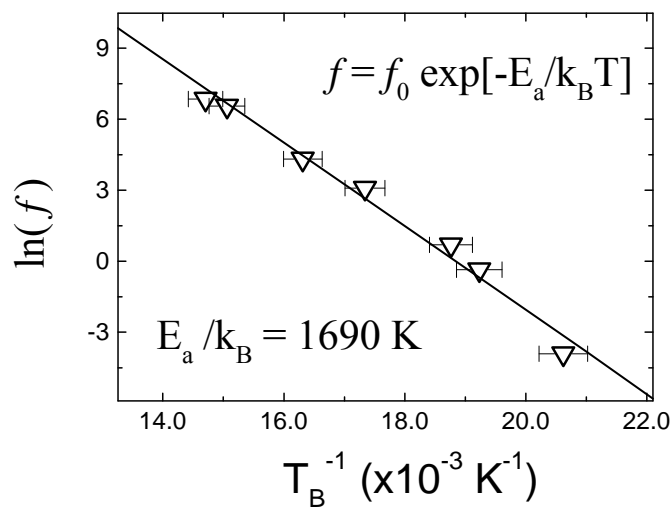

Figure 3. Semilog plot of the inverse blocking temperature $T_B^{-1}$ vs. driven frequency $f$ obtained from the imaginary component χ´´(T). Solid line is the best fit using Eq. (1) with $E_a/k_B$ = 1690 K.

Regarding the blocking/freezing process, the existence of particle interactions hinders the analysis of the single-particle magnetic anisotropy, since contributions from the neighbor magnetic dipoles to the local field can be even larger than the intrinsic (crystalline or shape) anisotropy. Additionally, surface effects have been invoked as a large source of magnetic anisotropy. However, from symmetry arguments, Bødker et al. [8] have demonstrated that a perfect spherical particle should have a zero net contribution from surface anisotropy. Since TEM images showed that the present Fe$_3$O$_4$ particles are nearly spherical, no major contribution from the surface should be expected. Consequently, the observed enhancement of the particle anisotropy should be mainly related to the effect of dipolar interactions rather than to a surface with larger anisotropy. In a previous work on Fe$_3$O$_4$ particles in a ferrofluid having similar mean diameter than our sample, [9] it was found that dipolar interactions are already noticeable for concentrations of ~2 % vol. of magnetic particles.

The larger blocking temperature observed at low fields in our sample further agrees with the presence of strong dipolar interactions due to the higher concentration of magnetic particles. Therefore, the value of $K_{eff}$ obtained for the present non-diluted system should contain the effect of particle-particle interactions.

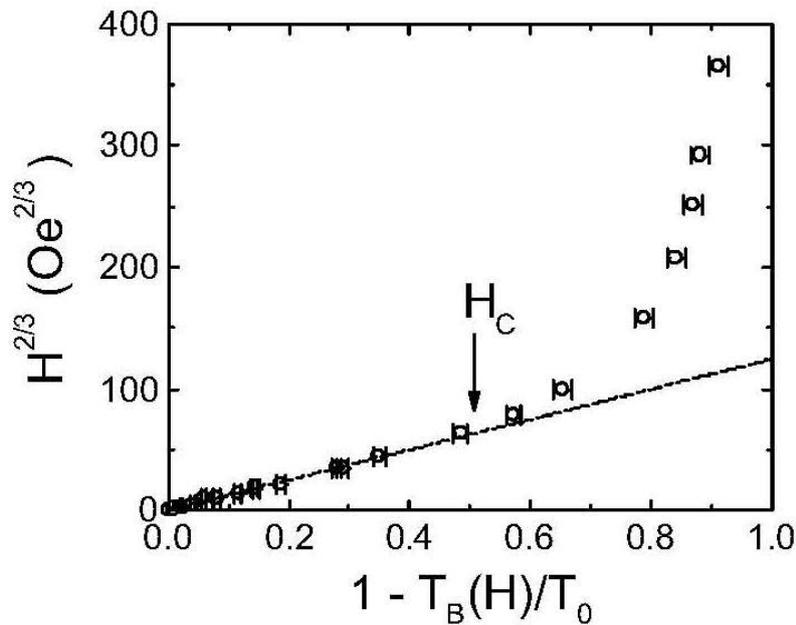

**Figure 4. Dependence of the blocking temperature $T_B$ on the applied field H (open circles). The solid line is the best fit using the power law $H^{2/3} = H_0 (1-T_B/T_0)$ for H < 550 Oe, as discussed in the text.**

To see whether these samples show spin-glass-like freezing, we followed the evolution of $T_B$, as obtained from the maximum of the ZFC magnetization curves measured at different applied fields. Figure 4 shows that for H = 550 Oe, this evolution can be described by a power law dependence of the blocking temperature, i.e. $H^{2/3} \sim (1- T_B/T_0)$, where $T_0$ represents the blocking temperature at zero field. This dependence corresponds to the Almeida-Thouless (AT) line observed in spin glasses, but also obtained from numerical calculation of the Brown equation for the relaxation time t in a pure SPM system, under the assumption of $E_a$ = constant, where $E_a$ is the single-particle energy barrier.[11] The latter model is clearly more appropriate given the nanostructured nature of our samples, but also because the fields involved here are several orders of magnitude larger than for spin glasses. As seen from figure 4, the $H^{2/3}$ behavior is observed up to $H_C \approx 550$ Oe, where the curve bends over to a different $H^\lambda$ dependence for fields H > 2 kOe. The two different regimes can be explained qualitatively by considering the effect of the applied field on the SPM relaxation time, calculated by Brown [12] for a high energy barrier approximation (in the limit $\mu H_{app} \ll E_a$, where m is the particle moment), and after numerically calculated for $\mu H_{app} \sim E_a$ [11]. When the applied field is much lower than a certain value $H_C$ that satisfies $\mu H_C \ll E_a$, the equations can be linearized resulting in a $H^2 \sim (1-T_B/T_0)$ dependence. [11]. From this value of the crossover field $H_C$, we estimated the average height of the energy barriers $E_a = 6.2 \times 10^{-22}$ J. It has been demonstrated [10] that for weakly interacting $Fe_3O_4$ particles these two field regimes are separated by $H_C \sim 150$ Oe, and for the high field regime the values of $T_B$ do not depend on the particle interactions. In the present case, the higher concentration results in larger $T_B$ values, and accordingly we calculated $H_C \sim$ 450-600 Oe, in good agreement with the value of 550 Oe found from the linear fit (fig. 4).

**Summary**

We have studied the magnetic properties of nearly spherical magnetite particles as a function of temperature and field, finding that the effects of dipolar interactions are dominant at low fields. Both magnetization and ac susceptibility data are well described by a thermally activated, SPM model. The frequency dependence of the susceptibility maxima provides a value of the single-particle energy barriers much larger that the expected from magnetocrystalline anisotropy, due to the influence of interparticle interactions. Two different regimes have been found in the H-T plane, both explained by the high and low field solutions of the Brown equation for the relaxation time of single particles. The crossover from the low to the high field dependence gives an estimation of the average energy barrier of $E_a = 6.21 \times 10^{-22}$ J.

**Acknowledgements**

This work was done under financial support from Brazilian agencies FAPESP and CNPq.